\documentstyle[12pt]{article}

\begin{document}                % End of preamble and beginning of text

\baselineskip=20pt
\newcommand{\beq}{\begin{equation}}
\newcommand{\eeq}{\end{equation}}
\newcommand{\beqa}{\begin{eqnarray}}
\newcommand{\eeqa}{\end{eqnarray}}
\newcommand{\bc}{\begin{center}}
\newcommand{\ec}{\end{center}}
\newcommand{\n}{\newline}
\newcommand{\ra}{\,\rangle}
\newcommand{\la}{\langle\,}
\newcommand{\nn}{\nonumber\\}
\newcommand{\ov}{\overline}
\newcommand{\bi}{\bibitem}
\newcommand{\cl}{{\cal C}}
\newcommand{\gl}{{\cal G}}
\newcommand{\f}{{\bar f}}
\newcommand{\x}{({\bf x}_o;t_o)}
\newcommand{\k}{({\bf k_o};t_o)}
\newcommand{\xp}{({\bf x}_o';t_o')}
\newcommand{\kp}{({\bf k_o}';t_o)}
\newcommand{\th}{\theta({\bf x}_o;t_o)}
\newcommand{\thp}{\theta({\bf x}_o';t_o')}
\newcommand{\ff}{\phi({\bf x};t)}
\newcommand{\ffp}{\phi({\bf x}';t')}
\newcommand{\xm}{(x_o;t_o,t_o')}
\newcommand{\xr}{({\bf x};t)}
\newcommand{\kr}{({\bf k};t)}
\newcommand{\xpr}{({\bf x'};t')}
\newcommand{\kpr}{({\bf k'};t)}
\newcommand{\xmr}{(x;t,t')}
\newcommand{\thr}{\theta({\bf x};t)}
\newcommand{\thpr}{\theta({\bf x}';t')}
\newcommand{\pp}{(2 \pi)^2}
\newcommand{\G}{\Gamma}
\renewcommand{\theequation}{\thesection.\arabic{equation}}

\pagestyle{empty}
\begin{flushright}
Univ. di Roma I preprint {\bf N.1028}

{\bf cond-mat }
\end{flushright}

\vspace{1cm}

\bc
{\Large {
{\bf Off equilibrium dynamics and aging in unfrustrated systems}
}}
\ec
\vspace{.3cm}
\bc
{\large {
L. F. Cugliandolo,
J. Kurchan and
G. Parisi
}}
\ec
\vspace{.1cm}
\bc
{\large {
Dipartimento di Fisica,
}}
\linebreak
{\large {
Universit\`a di Roma I,
{\it La Sapienza},
}}
\linebreak
{\large {
I-00185 Roma, Italy
}}
\linebreak
{\large {
INFN Sezione di Roma I, Roma, Italy
}}
\linebreak
\ec
\vspace{.2cm}
\bc
{\large {
\today
}}
\ec
\vspace{0.05cm}

\newpage

\bc
{\large {\bf Abstract}}
\ec

\vspace{.4cm}

We analyse the Langevin dynamics of the random
walk, the scalar field, the X-Y model and
the spinoidal decomposition. We study the deviations
from the equilibrium dynamics theorems (FDT and homogeneity),
the asymptotic behaviour of the systems and  the
aging phenomena. We compare
the results with the dynamical behaviour of (random)
spin-glass mean-field models.

\vspace{3cm}

\newpage

\pagestyle{plain}
\setcounter{page}{1}
\setcounter{footnote}{1}

\vspace{1cm}
\section{Introduction}
\setcounter{equation}{0}
\vspace{1cm}

 Spin-glasses and other disordered systems have `critical'
dynamics throughout their low-temperature phase.
 Their most striking dynamical effect is that of aging:
they do not reach thermal equilibrium after very long times and
experiments are performed out of equilibrium
showing a dependence on the history of the system
\cite{Lusvnobe,Vihaoc}.

These phenomena have been studied with numerical simulations
\cite{Anmasv,Ri1,Cukuri,Mapa},
several phenomenological models have been proposed \cite{pheno,Bo1},
and analytical techniques have
applied to mean-field systems \cite{Cuku,Cuku2,Frme}.

It has been recently proposed \cite{Cuku}
that mean-field spin-glasses do not to reach
a situation of dynamical equilibrium ({\it i.e.}
homogeinity in time and the fluctuation-dissipation
theorem (FDT) are violated) even after infinitely long times.

In other systems an intermediate situation known as {\em interrupted aging}
occurs in which non-equilibrium effects tend to dissapear, but very
slowly as compared to the relaxation of ordinary non-critical systems
({\it e.g.}
paramagnets, ferromagnets) \cite{Bo2,Ri2}.

The persistence of out-of equilibrium effects
after very long times, and in particular
the violation of the equilibrium theorems  is a feature
not restricted to disordered systems
such as spin-glasses. It is interesting in itself
to study the deviations from the equilibrium theorems in
simpler examples with  hamiltonians that are deterministic (non random), and
even not disordered or frustrated.

The aim of this note is to analyse the Langevin dynamics of
some such simple examples: the random walk, the free scalar field
and the X-Y model,  and the spinoidal decomposition of a ferromagnetic
 Ising-like system.
 In each case we shall
study the deviations from the equilibrium theorems and we shall
analyse the long-time behaviour of the correlation and  response
functions,
and the total response to a constant perturbation applied during a
finite time-interval (the equivalent of the
`thermoremanent magnetization' in spin-glass experiments).

The organisation of the paper is the following.
In section 2 we present some general remarks on the FDT
and its possible generalization. In section 3 we analize the simplest
non equilibrium model, {\it i.e.} the random walk. In section 4 we consider the
case of the $D$-dimensional free scalar field theory.
In section 5 we study the
dynamics of the X-Y model at low temperature in two dimensions.
Finally in section 6 we consider the dynamics of the spinoidal
decomposition for the usual ferromagnetic Ising case. Our conclusions are
presented in section 7.

\vspace{1cm}
\section{The generalized fluctuation dissipation relation}
\setcounter{equation}{0}
\vspace{1cm}

Let us consider a system which has been quenched
from high temperature at time $t=0$. The
auto-correlation function $C(t,t')$ among a local quantity
$O$ at two subsequent times $t'$ and $t$ is
\beq
C(t,t') = \la O(t) O(t') \ra
\; .
\eeq
Hereafter $\langle \;\cdot\;\rangle$
represents the mean over the thermal noise.

For large $t$ and fixed $t-t'$, in an equilibrium dynamics situation
the auto-correlation function behaves as
\beq
C(t,t') =C(t-t')
\; ,
\eeq
{\it i.e.} it is homogeneous in time.

The response function to a pertubation
is defined as the variation
of the quantity $\la O(t) \ra$ with respect to a perturbation applied at
time $t'$.
More precisely, if we consider the perturbed Hamiltonian
\beq
H= H_o +\int dt \ h(t)O(t)
\; ,
\eeq
the response function is defined as
\beq
R(t,t')={\delta \la O(t) \ra \over \delta h(t')}
\eeq
and, because of causality, it equals zero if $t' > t$.
The response function is not independent of the correlation if the system
is in equilibrium.
Indeed, it is related to the correlation function by the
celebrated fluctuation-dissipation theorem (FDT):
\beq
R(t,t') = \beta \, \theta(t-t') \, \frac{\partial C(t,t')}{\partial t'}
\; ,
\label{fdt}
\eeq
and it is also homogeneous in time $R(t,t') = R(t-t')$.

However, if the system is out of equilibrium neither homogeneity nor
the FDT (\ref{fdt}) hold.
The generalised relation between response and
correlation functions can be written as
\beq
R(t,t') = \beta \theta(t-t') \, X(t,t') \, \frac{\partial C(t,t')}{\partial t'}
\label{fdtgen}
\eeq
with $X$ a function of both times $t'$ and $t$
that characterizes the approach to equilibrium.

\vspace{1cm}

The FDT and its violation can be partially
understood from the following considerations.
Let us consider a system described a variable $y(t)$ which satisfies the
Langevin equation
\beq
\frac{d}{d t} y(t) = - F[y](t) + \eta(t)
\label{lang0}
\eeq
where  $\eta$ is a Gaussian random noise with zero mean and correlation
\beq
\la \eta(t) \, \eta(t') \ra = 2 T \, \delta (t-t')
\; ,
\label{eta}
\eeq
$T$ being the temperature.

Taking $t>t'$ for definiteness, the equation of motion (\ref{lang0}) implies
\beq
({\partial \over \partial t'} - {\partial \over \partial t}) \, C(t,t')
= 2 T R(t,t') + A(t,t')
\label{giorgio1}
\eeq
where we have used $\la y(t) \, \eta(t') \ra = 2T \, R(t,t')$ and
\beq
A(t,t') \equiv \la F[y](t) \, y(t') - F[y](t') \, y(t)\ra
\; .
\eeq

At equilibrium the correlation functions satisfy
$ \la B(t) D(t') \ra = \la B(t') D(t) \ra$, if $B(t)$ and
$D(t')$ are any two functions of $y(t)$. This is
a consequence of the time
reversal symmetry. Hence the asymmetry $A$ vanishes and
the fluctuation-dissipation
theorem may be recovered by using the invariance under translations in
time of the correlation functions at equilibrium:
\beqa
\begin{array}{rclcrcl}
C(t,t') &=& C(t-t')
&
\Rightarrow
&
\left(
{\partial \over \partial t'} +{\partial \over \partial t}
\right) \,
C(t,t') &=& 0
\label{hom}
\end{array}
\eeqa
and
\beq
R(t,t') = \beta \, \frac{\partial C(t,t')}{\partial t'}
\; .
\eeq

In the off-equilibrium situation the homogeneity in time (eq.(\ref{hom}))
is not valid and the asymmetry $A$ may be present. Eq. (\ref{fdt})
is not valid in general and the generalisation (\ref{fdtgen})
must be considered.

\vspace{1cm}

In mean-field spin-glass models the auto-correlation and response functions
are defined as $C(t,t')$ $=$ $(1/N)\sum_{i=1}^N \la s_i(t) s_i(t') \ra$ and
\linebreak
$R(t,t')$ $=$ $(1/N) \sum_{i=1}^N \delta \la s_i(t) \ra / \delta h_i(t') $,
respectively.
In the analysis of the asymptotic dynamics
\footnote{Asymptotic means $t , t' \rightarrow \infty$ {\em after}
$N \rightarrow
\infty$}
presented in Ref. \cite{Cuku2} (see also \cite{Ho}) it has been
proposed that,
for long enough times and  small time differences,
$t,t' \rightarrow \infty$ and $(t-t')/t << 1$,
$X=1$ and FDT is satisfied, while
for long enough times and big time differences,
$t,t' \rightarrow \infty$ and $(t-t')/t \sim O(1)$,
the function $X$ depends on the
times only through
\footnote{ This result is expected to hold for large times for systems
with {\em finite} susceptibility.}
 the correlation function $C(t,t')$, {\it i.e.}
\beq
R(t,t') = \beta \theta(t-t') \, X[ C(t,t') ] \,
\frac{\partial C(t,t')}{\partial t'}
\; .
\label{fdtasym}
\eeq
A self-consistent  asymptotic solution for the  mean-field out of equilibrium
dynamics within this assumption has been found both for the $p$-spin
spherical
and the Sherrington-Kirkpatrick models
(Refs. \cite{Cuku},\cite{Cuku2}.

In the following sections
we shall investigate the behaviour of the function
$X$ for various (non random) models and we shall compare the
results with expression (\ref{fdtasym}) at long times.

\vspace{1cm}

\subsection{Scalings and aging phenomena}
\setcounter{equation}{0}
\renewcommand{\theequation}{\thesubsection.\arabic{equation}}

\vspace{1cm}

Another interesting problem is to study the
scaling properties of the
correlation, response and $X$ functions, and
the response of the system to a constant perturbation
applied during a finite period $[0,t_w]$.

For the $p$-spin spherical spin-glass model,
if $t$ and $t-t'$ are both large
one analytically finds
\beq
C(t,t') \propto \frac{h(t')}{h(t)}
\eeq
within the assumptions described above. The numerical solution
of the dynamical equations suggests that
\beq
C(t,t') = S(t'/t)
\; ,
\label{sca1}
\eeq
({\it i.e.} $h$ a power law). This is a new (non-homogeneous) scaling
\footnote{In the dynamics of other mean-field spin-glass models
more complicated scalings can be present.}.
This is the simplest scaling that captures
an essential feature of spin-glass phenomenology:
the aging effects, {\it i.e.} the explicit dependence
of the behaviour of the system on its history.
The scaling (\ref{sca1}) can be modified in many
ways to describe in more detail the results of simulations of
realistic models and
experiments.
Then, one sometimes assumes the slightly different form
\beq
C(t,t') = t^{-\delta} \, S(t'/t)
\; ,
\label{Cscaling}
\eeq
where $\delta$ is a small number, of the order of a few percent
in spin-glass models \cite{Vihaoc,Bo1,Ri1}. The factor $t^{-\delta}$
implies an interruption of aging for large $t$ ($O($few years$)$).

The generalised FDT relation (\ref{fdtgen}) can be written as
\beq
R(t,t')= \beta \, \Theta(t-t') \,
X[t^{-\delta}, t'/t] \; {\partial \over \partial t'}C(t,t'),
\label{RvsdC}
\eeq
If we substitute the scaling form
(\ref{Cscaling})
for the correlation function in eq.(\ref{giorgio1})
and we assume that a similar form is valid for the asymmetry,
we find that the response
function scales as $C/t'$ or $\partial C/ \partial t'$.

If instead we assume that
the asymmetry $A$ is zero, as will be the case below, we find
\beq
X(t,t')
=
X(\lambda)
=
\frac{1}{2} \left[ 1 + \lambda + \delta \, \frac{S(\lambda)}{S'(\lambda)}
\right]
\label{xz}
\eeq
with $\lambda\equiv t'/t$.

\vspace{.5cm}

In the typical aging experiments \cite{Lusvnobe,Vihaoc}
one measures the `thermoremanent
magnetization', {\it i.e.}
the response of the system to a constant magnetic field $h$ applied
during the interval $[0,t_w]$, at constant temperature.
$t_w$ is interpreted as a `waiting time'.
In a general dynamical system described by
the Langevin equation (\ref{lang0}) the equivalent of the
thermoremanent magnetization is
\beq
\chi_{t_w} (t)
=
\int_o^{t_w} dt' R(t,t')
\; .
\label{ther}
\eeq
Aging experiments show
that $\chi_{t_w}(t)= m_{t_w}(t)$
depends non-trivially on $t$ and $t_w$ \cite{Lusvnobe,Vihaoc}.

\vspace{1cm}
\section{The Random Walk}
\setcounter{equation}{0}
\renewcommand{\theequation}{\thesection.\arabic{equation}}
\vspace{1cm}

The simplest example of a dynamical system that does not reach equilibrium is
the random walk \cite{Vi}.
In the continuum limit the quantity $y(t)$ satisfies the very
simple differential equation
\beq
\frac{d}{d t} y(t) =  \eta(t) ,
\eeq
with $\eta$ a Gaussian noise with variance given by eq.(\ref{eta}).

It is easy to check that the correlation function
$C(t,t') = \la y(t) \, y(t') \ra$ and the
response function $R(t,t') = \delta \la y(t) \ra / \delta h(t') =$
$(\beta / 2) \, \la y(t) \eta(t') \ra$
are given by
\beqa
C(t,t') &=& 2 T \ \min(t,t') \; , \\
R(t,t') &=& \theta(t-t')  \; .
\eeqa
Hence, the relation (\ref{fdtgen}) is satisfied with
\beq
X(t,t') = 1/2
\; ,
\eeq
$\forall t,t'$,
a constant function but different from the usual FDT result, $X=1$,
the system never reaches equilibrium.

Indeed one finds that the
scaling form (\ref{Cscaling})
for the correlation is satisfied with
$S(\lambda)=\lambda$ for $\lambda < 1$,
but with a big value for $\delta$, $\delta = -1$.
Inserting this scaling in
eq.(\ref{xz}) we also obtain $X=1 /2$, as expected since in the random
walk problem the force $F$ and the asymmetry $A$ are zero.
However, the scalings for the correlation and the total response
are quite different from those observed in spin-glasses.
In terms of the
`waiting time' $t_w$ and $\tau \equiv t- t_w$,
$C(\tau + t_w, t_w) = 2T \, t_w$ and
$\chi_{t_w}(\tau + t_w)  = t_w$ ({\it cfr.} eq. (\ref{ther})).
Both expressions are independent of $\tau$ but depend explicitly
on $t_w$.

This example may seem trivial, but it captures the essence of the phenomenon
that will be described in the rest of this note.

\vspace{1cm}
\section{Free Gaussian Fields}
\renewcommand{\theequation}{\thesection.\arabic{equation}}
\setcounter{equation}{0}
\vspace{1cm}

In this section we study the behaviour of a simple free scalar
field $\phi\xr$.
The Hamiltonian is quadratic in the field and in dimensions $D$
it reads
\beq
H = \frac{1}{2} \int \, d^D {\bf x_o} \;
\left[
(\nabla \phi)^2 +  m_o^2 \, \phi^2
\right]
\eeq
where $m_o$ is the mass of the field
(see {\it e.g.} Ref.\cite{Zi}).

The relaxational dynamics is given by the Langevin equation
\beq
\frac{\partial}{\partial t_o} \phi \x
=
 \Delta \phi \x - m_o^2 \, \phi \x
+ \eta \x
\; .
\label{langphi}
\eeq
$\eta\x$ is a Gaussian noise ($\eta\k$  its Fourier transform)
with zero mean and correlations
\beqa
\begin{array}{rcl}
\la \eta\x \; \eta\xp \ra
&=&
2T \;
\exp
\left(
-\frac{x_o^2 \Lambda^2}{4}
\right)
\;
\delta(t_o-t_o')
\nn
\la \eta\k \; \eta\kp \ra
&=&
2T \; (2 \pi)^D \;
\exp
\left(
-\frac{k_o^2}{ \Lambda^2}
\right)
\;
\delta^D({\bf k_o} + {\bf k_o'})
\;
\delta(t_o-t_o')
\nonumber
\end{array}
\label{etascalar}
\eeqa
$x_o^2 \equiv |{\bf x}_o - {\bf x}'_o|^2$ and $k_o^2 \equiv |{\bf k_o}|^2$.
We have introduced a spatial correlation over a typical lenght
$1/\Lambda$ to simulate the lattice spacing. (This serves
to cure some large $k$ pathologies.)

Taking $\phi({\bf x_o}, 0) = 0$ as the initial condition,
the solution to the dynamical equation (\ref{langphi})
for each noise realisation is
\beq
\phi \x
=
\int \frac{d^D {\bf k_o}}{(2\pi)^D} \;
e^{i {\bf k_o} {\bf x_o}}
\;
\int_o^{t_o}
d\tau \; e^{-(k_o^2 + m_o^2) (t_o-\tau)} \; \eta({\bf k_o},\tau)
\; .
\label{phiscalar}
\eeq

\vspace{.5cm}

Since we are dealing with a field, the correlation,
response and $X$ functions depend on space-time coordinates.
A standard calculation for the correlation function
$C_o({\bf x_o}, {\bf x'_o}; t_o,t_o') = \la \phi \x \, \phi \xp \ra$
gives
\beqa
C_o({\bf x_o}, {\bf x_o'}; t_o,t_o')
&=&
T \int \frac{d^D {\bf k_o}}{(2 \pi)^D}
\frac{1}{k_o^2 + m_o^2}
e^{i {\bf k_o}({\bf x_o}-{\bf x_o'})} e^{-\frac{k_o^2}{\Lambda^2}} \;
\nonumber \\
& &
\left( e^{- (k_o^2 + m_o^2) (t_o-t_o')} - e^{-(k_o^2 + m_o^2)
(t_o+t_o')} \right)
\; .
\label{Cscalar}
\eeqa

The response function $R({\bf x_o}, {\bf x_o}';t_o,t_o') =
\partial \la \phi\x \ra/ \partial h \xp$
is given by
\beqa
R_o({\bf x_o}, {\bf x_o}';t_o,t_o')
&=&
\int \frac{d^D {\bf k_o}}{(2\pi)^D} \;
e^{i {\bf k_o}({\bf x_o}-{\bf x_o'})} e^{-\frac{k_o^2}{\Lambda^2}}
\;
e^{- (k_o^2+m_o^2) (t_o-t_o')}
\nn
&=&
\frac{1}{(4\pi)^{D/2}}
\;
\frac{e^{-m_o^2 (t_o-t_o')}}
     {(t_o-t_o'+ \frac{1}{\Lambda^2})^{D/2}}
\;
e^{-\frac{x_o^2}{4(t_o-t_o'+\frac{1}{\Lambda^2})}}
\; .
\label{Rscalar}
\eeqa
Here and in what follows we take unprimed times bigger than primed times
and we omit the theta functions.

The preceeding formul\ae $\;$
suggest to measure space, time and mass in appropriate
lattice units
\beqa
t &\equiv& \Lambda^2 t_o
\; ,
\\
{\bf x} &\equiv& \Lambda {\bf x}_o
\; ,
\\
m  &\equiv& m_o/\Lambda
\eeqa
and to rescale the correlation and response functions
$C \equiv \Lambda^{(2-D)} C_o $, and
$R \equiv (1/\Lambda^D) R_o$. Note that there is no
rescaling of fields and correlations in $D=2$.

The function $X$
that measures the departure from FDT reads, in terms of
the rescaled coordinates:
\beqa
X({\bf x}, {\bf x}'; t,t')
&\equiv&
T
\frac
   {R({\bf x}, {\bf x}';t,t')}
   {\frac
           {\partial  C({\bf x}, {\bf x'}; t,t')}
           {\partial t'}
   }
\nn
&=&
\left[ 1+ \left( \frac{t-t'+1}{t+t'+1} \right)^{D/2}
\exp \left\{
-2m^2t'+\frac{x^2 t'}{2[(t+1)^2-t'^2]}
\right\}
\right]^{-1}
\nn
\label{Xscalar}
\eeqa

%%%%%%%%%%%%%%%%%%%%%%%%%%%%%%%%%%%%%%%%%%%%%%%%%%%%%%%%%%%%%%%%%

\vspace{1cm}
\subsection{Large-times behaviour}
\setcounter{equation}{0}
\renewcommand{\theequation}{\thesubsection.\arabic{equation}}
\vspace{1cm}

%%%%%%%%%%%%%%%%%%%%%%%%%%%%%%%%%%%%%%%%%%%%%%%%%%%%%%%%%%%%%%%%%

Consider first the massive case. We have a time scale given by:
\beq
t_{eq} \sim m^{-1/2}
\; .
\eeq
For any $x$ fixed and
any two times $t,t'>>t_{eq}$ we have that $X=1$. This identifies
$t_{eq}$ as an `equilibration' time
\footnote{Note however that if $x$ is of order $\sqrt{t'}$ or larger
then $X$ can be smaller than $1$, even zero for small time differences.
We shall not consider
such diverging distances in the rest of the section.}.

In the massless case  $m=0$ the equilibration time diverges and we
have a more interesting situation. Let us concentrate in this case.
For fixed $x$ and large times $t,t'$, we have
\beq
X(x;t,t')
=
X(x;\lambda)
=
\frac{1}{1+ \left( \frac{1-\lambda}{1+\lambda} \right)^{D/2} }
\; ,
\label{xlambda}
\eeq
with $\lambda = t'/t$.
Hence $X$ is non-trivial and FDT is violated, even for very long
times.

If $\lambda \rightarrow 0$ then
\beq
X(x;\lambda) \rightarrow 1/2
\; .
\eeq

If $\lambda \rightarrow 1$ and $D\neq 0$, then
\beq
X(x;\lambda) \rightarrow 1
\; ,
\eeq
and we recover FDT. $\lambda=1$ corresponds to times
$t,t'$ satisfying $(t-t')/t <<1$, {\it i.e.} small
time differences.

If we put $D=0$ we recover $X=1/2$ for all times,
the result for the random walk.

%%%%%%%%%%%%%%%%%%%%%%%%%%%%%%%%%%%%%%%%%%%%%%%%%%%%%%%%%%%%%%%%%%%%%%

\vspace{1cm}
\subsection{Scalings}
\renewcommand{\theequation}{\thesubsection.\arabic{equation}}
\vspace{1cm}

%%%%%%%%%%%%%%%%%%%%%%%%%%%%%%%%%%%%%%%%%%%%%%%%%%%%%%%%%%%%%%%%%%%%%%%

We now present the scalings.
Since the massless scalar field turned out to be more interesting
we shall concentrate in this model.
If $m = 0$ the explicit computation of
the integrals in eq. (\ref{Cscalar}) gives
\beq
C({\bf x}, {\bf x'}; t,t')
=
T \frac{1}{\pi^{D/2}}
\frac{x^{2-D}}{4}
\Gamma
\left[
\frac{D}{2}-1; \frac{x^2}{4(t+t'+1)}, \frac{x^2}{4(t-t'+1)}
\right]
\; ,
\label{Cscalar2}
\eeq
where
$\G[n;a,b]$ is the generalised incomplete Gamma function
\beq
\G \left[ n;a,b \right] \equiv \int_a^b dz \; z^{n-1} e^{-z}
\; .
\eeq

For equal space points $x=0$ eq.(\ref{Cscalar2}) reduces to
\beq
C(0;t,t') =
T \;
\frac{1}{(4\pi)^{D/2}} \;
\frac{1}{1-\frac{D}{2}}
\left[
(t+t'+1)^{1-D/2} - (t-t'+1)^{1-D/2}
\right]
\; .
\label{c1}
\eeq
For long times and $\lambda < 1$, {\it i.e.} big time
differences $(t-t')/t \sim O(1)$,
this expression satisfies the scaling law (\ref{Cscaling}) with
$\delta=D/2 -1$ and
\beq
S(\lambda)
=
T \;
\frac{1}{(4\pi)^{D/2} \, (1-\frac{D}{2})}
(|1 - \lambda|^{1-D/2} + |1 + \lambda|^{1-D/2})
\; .
\eeq
Hence, in this time scale the function $X$ (eq. (\ref{xlambda}))
can be written
as $X(0;t,\lambda) = X[t^\delta C]$ and in particular
$X(0;t,\lambda) = X(C)$ for $D=2$.

\vspace{1cm}

Considering again the general model, the total response (\ref{ther})
reads
\beqa
\chi_{t_w}(t)
&=&
\int d^D{\bf x} \int_o^{t_w} dt''
\frac{1}{(4\pi)^{D/2}}
\frac{e^{-m^2(t-t'') -\frac{x^2}{4(t-t''+1)}}}{(t-t''+1)^{D/2}}
\nn
&=&
\frac{1}{m^2} e^{-m^2 (t-t_w)}
\left[ 1 - e^{-m^2 t_w} \right]
\;
\eeqa
and for large $t_w$, $t_w >> t_{eq}$ it reduces to
\beq
\chi_{t_w}(\tau + t_w)
=
\frac{1}{m^2} e^{-m^2 \tau}
\; ,
\eeq
the typical relaxation in a system that has equilibrated;
{\it i.e.} $\chi_{t_w}(\tau)$ depends only on $\tau = t - t_w$ and
no aging is present.

Instead, in the massless limit
\beq
\chi_{t_w}(\tau + t_w)
=
t_w
\eeq
which shows a dependence on the history for all $t_w$,
although a rather unusual one.

\vspace{1cm}

The learned reader will notice that in the massless
case the Hamiltonian is invariant under the transformation
\beq
\phi(x) \to \phi(x)+constant.
\eeq
The correlation functions are not invariant under this transformation and
therefore the symmetry is spontaneously broken. The slow approach to
equilibrium is a reflection in the time domain of the Goldstone boson arising
from the spontaneous breaking of the symmetry.
In the next section we shall see a case where the symmetry group is
implemented in a non linear way.

\vspace{1cm}

These results are in agreement with the
general formulae discussed in Section 2 when the
asymmetry is neglected
({\it cfr.} eq. (\ref{xz})). Indeed the
asymmetry is zero, because the force  $F$ is linear in the field $\phi$.

%%%%%%%%%%%%%%%%%%%%%%%%%%%%%%%%%%%%%%%%%%%%%%%%%%%%%%%%%%%%%%%%%%%%%%%
\vspace{1cm}
\section{The relaxational dynamics of the XY model}
\label{I}
\setcounter{equation}{0}
\renewcommand{\theequation}{\thesection.\arabic{equation}}
\vspace{1cm}
%%%%%%%%%%%%%%%%%%%%%%%%%%%%%%%%%%%%%%%%%%%%%%%%%%%%%%%%%%%%%%%%%%%%%%%

The Hamiltonian of the $O(2)$ non-linear $\sigma$ model can be written in
terms of the angular variable $\theta$ defined through
${\bf S}({\bf x}_o) = (\cos \theta({\bf x}_o), \sin \theta({\bf x}_o))$.
In two dimensions it reads
\beq
H = \frac{1}{2} \int \, d^2 {\bf x}_o \; (\nabla \th)^2
\eeq
(see {\it e.g.} Ref. \cite{Zi}).

The relaxational dynamics is given by the Langevin equation
\beq
\frac{\partial}{\partial t_o} \th
=
- \frac{\delta H}{\delta \th}
+ \eta\x
\; ,
\label{lang}
\eeq
with $\eta\x$ as in eq. (\ref{etascalar}).

We  consider low temperatures such that
vortices can be neglected and
therefore we do not see the Kosterlitz-Thouless transition.

The solution to the dynamical equation (\ref{lang})
for each noise realisation is that of the massless
scalar field problem in $D=2$, {\it cfr.} eq.(\ref{phiscalar}).

The angle-angle correlation
$\tilde C({\bf x}_o, {\bf x}_o'; t_o, t_o')$
$\equiv$
$\la \theta\xp \; \theta\x \ra$ is given by
\beq
\tilde C\xm
=
\frac{T}{4 \pi}
\;
\G
\left[ 0;
\frac{\Lambda^2 x_o^2}{4(1 + \Lambda^2 (t_o+t_o'))},
\frac{\Lambda^2 x_o^2}{4(1 + \Lambda^2 (t_o-t_o'))}
\right]
\label{cotilde}
\; ,
\eeq
and in particular, the time correlation between the angles at the same
space point ($x_o^2 = 0$) is
\beq
\tilde C(0;t_o,t_o')
=
\frac{T}{4 \pi} \log \left( \frac{1+ \Lambda^2 (t_o+t_o')}
{1+ \Lambda^2 (t_o-t_o')}
\right)
\; .
\eeq

The response to an external field $\tilde h \x$ acting like $\tilde h\x \;
\th$,
$\tilde R_o({\bf x}_o, {\bf x}_o'; t_o, t_o')
\equiv
\delta \la \theta\x \ra / \delta \tilde h \xp =
1/(2T) \,
\la \theta\x \; \eta\xp \ra$ is given by
\beq
\tilde R_o\xm
=
\frac{\Lambda^2}{4\pi}  \;
\frac{
      \exp \left(
                  -\frac{\Lambda^2 x_o^2}{4 (1 + \Lambda^2 (t_o-t'_o))}
           \right)}
     {1 + \Lambda^2 (t_o-t_o')}
\; .
\eeq

We now turn to calculating the physical quantities for which the
angular character of $\theta$ is essential.
We first calculate the `composite' correlation
\beqa
C\xm
&\equiv&
\la \sin \th \; \sin \thp \ra
\\
&=&
\exp
\left(
-\frac{1}{2}
\left( \tilde C(0;t_o,t_o) + \tilde C(0;t_o',t_o') \right)
\right)
\sinh \tilde C \xm
\nn
\label{compc}
\eeqa
and the associated response to
a transverse field $h \x$ acting like \newline
$h \x \sin \th$
\beq
R_o \xm \equiv \frac{\delta m\x}{\delta h\xp}
\eeq
where $m\x$ is the
transverse magnetisation $m\x = \la \sin \th \ra$.
The `composite' response can also be written in terms of the
angle-angle correlation $\tilde C$ and its associated response function
$\tilde R_o$:
\beqa
R_o \xm
&=&
\exp
\left(
-\frac{1}{2}
\left(
\tilde C(0;t_o,t_o) + \tilde C(0;t_o',t_o')
\right)
+
\tilde C\xm
\right)
\; \nn
& &
\times \; \tilde R_o \xm
\label{compg}
\; .
\eeqa

As in the previous section we now rescale space-time coordinates
as
\beqa
t &\equiv& \Lambda^2 t_o
\; ,
\\
{\bf x} &\equiv& \Lambda {\bf x}_o
\; ,
\eeqa
and rescale the  response function
$R \equiv (1/\Lambda^2) R_o$, $\tilde R \equiv (1/\Lambda^2) \tilde R_o$
(but neither the correlations nor the angles).

In terms of
the new coordinates we have
\beqa
\tilde C \xmr
&=&
\frac{T}{4\pi} \,
\G
\left[
0;
\frac{ x^2}{4(1 +  t + t')},
\frac{ x^2}{4(1 +  t - t')}
\right]
\label{tildec}
\; ,
\\
\tilde R \xmr
&=&
\frac{1}{4\pi} \,
\frac{1}{1+t-t'} \,
\exp \left( -\frac{x^2}{4(1+t-t')} \right)
\label{tildeg}
\; ,
\\
C\xmr
&=&
\left( (1 + 2  t) (1 + 2  t') \right)^{-\frac{T}{8\pi}}
\sinh \tilde C \xmr
\label{C}
\; ,
\\
R\xmr
&=&
\frac{1}{2}
\left( (1 + 2  t) (1 + 2  t') \right)^{-\frac{T}{8\pi}}
\tilde R \xmr
\exp \left( \tilde C \xmr \right)
\label{G}
\; .
\eeqa

{}From eqs. (\ref{C}) and (\ref{G}),
the function $X$ reads
\beq
X \xmr =
\frac{ 2 \tilde R \xmr}
     {A_- \xmr  + A_+ \xmr \exp (-2 \tilde C \xmr)}
\eeq
with
\beqa
A_{\pm} \xmr &\equiv& \frac{\partial \tilde C \xmr}{\partial t'}
\pm \frac{1}{2} \frac{\partial \tilde C(0; t',t')}{\partial t'}
\nn
&=&
\frac{T}{4\pi} \;
\left[ \,
\frac{ \exp \left(- \frac{ x^2}{4(1 +  t-t')}
\right) }
     { 1 +  t-t'}
+
\frac{ \exp \left(- \frac{ x^2}{4(1 +  t+t')}
\right) }
     { 1 +  t + t'}
\pm
\frac{1}{1+  2t'}
\,
\right]
\; .
\nn
& &
\eeqa

%%%%%%%%%%%%%%%%%%%%%%%%%%%%%%%%%%%%%%%%%%%%%%%%%%%%%%%%%%%%%%%%%%%%%

\vspace{1cm}
\subsection{Large-times behaviour}
\label{II}
\setcounter{equation}{0}
\renewcommand{\theequation}{\thesubsection.\arabic{equation}}
\vspace{1cm}

%%%%%%%%%%%%%%%%%%%%%%%%%%%%%%%%%%%%%%%%%%%%%%%%%%%%%%%%%%%%%%%%%%%%%

In this subsection we consider the large times limit,
$t$ and $t'$ large ($t>t'$). In this limit
the function $X$ is
\beq
\lim_{t\rightarrow\infty} X \xmr
=
\lim_{t\rightarrow\infty}
\frac{1}{1 + \exp \left( -2 \tilde C \xmr \right)}
\label{x}
\eeq
with $\tilde C \xmr $ given by eq. (\ref{tildec}).
We shall analyse the function $X$ and the correlation $C$ in different
regions determined by the space and time separations $x$ and $t-t'$.

\vspace{.5cm}
\noindent {\it Equal Times}
\vspace{.5cm}

\noindent We first consider the correlation and response functions at equal
times $t=t'\gg 1$.
We consider separately the cases $x=0$ (local values) and
$x\gg 1$ (many `lattice spacings').

\vspace{.5cm}
\noindent {\it a.} $\;\;\;\;\;\;\;\;\;\;\;x=0$
\beq
C(0;t,t)=\frac{1}{2}
\eeq
which was to be expected, since the $O(2)$ symmetry is unbroken and
\linebreak $\la \sin^2 \theta \ra = 1/2$. We also have
\beq
X(0;t,t)=1
\eeq
{\it i.e.} the system evolves locally with an equilibrium dynamics.

\vspace{.5cm}
\noindent {\it b.} $\;\;\;\;\;\;\;\;\;\;\;x^2\gg 1$

\vspace{.5cm}
\noindent {\it b.i} $\;\;\;\;\;\;\;\;\;\;\;t\gg x^2\gg 1$
\beq
\lim_{t\rightarrow\infty} X (x;t,t) = 1
\eeq
and
\beq
C(x;t,t) \rightarrow C_{static}(x) \simeq
x^{-\frac{T}{2\pi}}
\; .
\eeq

\vspace{.5cm}
\noindent {\it b.ii} $\;\;\;\;\;\;\;\;\;\;\;x^2\gg t\gg 1$
\beqa
\lim_{t\rightarrow\infty} X(x;t,t) &=& \frac{1}{2} \; ,
\nn
C (x;t,t) &\rightarrow& 0 \; .
\eeqa

We conclude that if we take a snapshot of the system at a large time $t$,
within a range of length $\simeq t^{1/2}$ the system
seems equilibrated in the sense that the correlations coincide with the static
ones \cite{Zi}
and the response satisfies FDT. Well outside that range the angles
are uncorrelated
and $X=1/2$, as in a random walk.
In the following we shall see in more detail the nature of this
`equilibration'.

\vspace{.5cm}
\noindent {\it Different Times}
\vspace{.5cm}

\noindent
We here consider different times, {\it i.e.} $t-t' \neq 0$ and we again
analyse separately the cases $x=0$ and $x \gg 1$.

\vspace{.5cm}
\noindent {\it a.} $x=0$
\beq
\lim_{t\rightarrow\infty} X(0;t,t')
=
\frac{1}
{1 +
\left( \frac{t+t'}{1+t-t'} \right)^{-\frac{T}{2\pi}} }
\eeq
and the correlation $C(0;t,t')$ reads
\beq
C(0;t,t')  \sim
\frac{1}{2} (4t t')^{-\frac{T}{8\pi}}
\left(
\left( \frac{t+t'}{1+t-t'} \right)^{\frac{T}{4\pi}} -
\left( \frac{t+t'}{1+t-t'} \right)^{-\frac{T}{4\pi}}
\right)
\label{cxxx}
\eeq

\vspace{.5cm}
\noindent {\it b.}
Consider now two points widely separated $x \gg 1$ but well
within the `range of equilibration' for these times, {\it i.e.}
\beq
t>t'\gg x^2\gg 1
\eeq
Two possibilities then arise:

\vspace{.5cm}
\noindent {\it b.i} $\;\;\;\;\;\;\;\;\;\;\;(t-t')/(t+t')\ll 1$

\noindent
($\lambda \rightarrow 1$.)
We reobtain the `equilibrium' situation $X=1$ and the correlation
function goes, asymptotically in $t-t'$, as
\beq
C \xmr \sim (t-t')^{-\frac{T}{4\pi}}
\; .
\eeq

\vspace{.5cm}
\noindent {\it b.ii} $\;\;\;\;\;\;\;\;\;\;\;(t-t')/(t+t')>0$

\noindent
($\lambda<1$.)

\beq
\lim_{t\rightarrow\infty} X \xmr = X(\lambda)
=
\frac{1}{1 + (\frac{1+ \lambda}{1 - \lambda})^{-\frac{T}{2\pi}}}
\; ,
\eeq
and the correlation function is
\beq
C(x;t,t') \simeq
\frac{(2t)^{-\frac{T}{4\pi}}}{2}
\lambda^{-\frac{T}{8\pi}}
\left[
\left( \frac{1+\lambda}{1-\lambda} \right)^{\frac{T}{4\pi}} -
\left( \frac{1-\lambda}{1+\lambda} \right)^{\frac{T}{4\pi}}
\right]
\eeq
({\it cfr.} eq. (\ref{cxxx})), and $X=t^{-T/(4\pi)} \tilde X[C]$.

\vspace{1cm}

For relatively small time differences with respect to the total time
and for  any two points well within a `domain' of equilibration
these results are still those of a system evolving as in equilibrium,
in other words $X=1$ and the correlation and response functions are
homogeneous in time.
However, even within a `domain' when
the time separation is large enough, the correlation and response  give
manifestly out of equilibrium results: $X$ has a non-trivial time
dependence and the correlation and response functions are not necessarily
homogeneous in time. We conclude
that one cannot picture these domains as regions in which
a true (lasting) equilibrium has been established. This has to be
contrasted with the behaviour of the massive scalar field, which
after a certain $t_{eq}$ and for fixed $x$ evolves as in equilibrium.

%%%%%%%%%%%%%%%%%%%%%%%%%%%%%%%%%%%%%%%%%%%%%%%%%%%%%%%%%%%%%%%%%%%%%%

\vspace{1cm}
\subsection{Remanent magnetization}
\label{III}
\renewcommand{\theequation}{\thesubsection.\arabic{equation}}
\setcounter{equation}{0}
\vspace{1cm}

%%%%%%%%%%%%%%%%%%%%%%%%%%%%%%%%%%%%%%%%%%%%%%%%%%%%%%%%%%%%%%%%%%%%%%%%%

Let us now turn to studying the behaviour of the system in a `thermoremanent
magnetization' experiment.

The total response to a constant, uniform  magnetic field $h\xmr$ applied from
$t'=0$ to $t'=t_w$ over the whole system is
\beq
\chi_{t_w}(t)
= \frac{1}{4}
(1+2t)^{-\frac{T}{8\pi}}
\int_o^{t_w} dt' \;
(1+2t')^{-\frac{T}{8\pi}}
\;\;
k \left( \frac{1+t-t'}{1+t+t'}, T \right)
\eeq
with
\beq
k ( w, T )
\equiv
\int_o^\infty du \; \exp \left( -\frac{u}{2} \right)
\exp \left( -\frac{T}{4\pi}
\G
\left[
0; \frac{u}{2}, \frac{u w}{2}
\right]
\right)
\; .
\eeq

Defining a function  (see Appendix A)
\beq
f(\lambda,T)
=
2^{-(1+\frac{T}{4\pi})}
\;
\lambda^{\frac{T}{4\pi}-1}
\int_{\frac{1-\lambda}{1+\lambda}}^1
\frac{dw}{(1+w)^2} \left( \frac{1-w}{1+w} \right)^{-\frac{T}{8\pi}}
k(w,T)
\; ,
\label{funct}
\eeq
we have that in the large $t$ limit for every $t_w$ and $t$
\beq
\chi_{t_w}(t)
=
t_w^{1-\frac{T}{4\pi}} f\left(\frac{t_w}{t}, T \right)
\; ,
\label{chifinal}
\eeq
and $f(\lambda_w,T)$ is an increasing function of $\lambda_w$, finite at
$\lambda_w=1$.
Asymptotically, for $t\gg t_w$
\beq
\chi_{t_w}(t)
\propto
t_w^{1-\frac{T}{8\pi}}  \; t^{-\frac{T}{8\pi}}
\label{chiasym}
\eeq

Several considerations are in order about this behaviour.
Throughout the low temperature phase $T < 4\pi$, and
 the susceptibility
diverges with $t_w$. This result was to be expected since
the static magnetization grows as $h^\gamma$ with $\gamma =
T/(8\pi -T)$ \cite{Zi}.

For finite times and small fields
the linear response theory holds, but becomes worse as
an approximation for larger times
and it fails completely at $t_w \rightarrow \infty$. This result is
reminiscent of what seems to happen in spin-glasses with the reaction of
the system when temperature is slightly changed: while experimentally
(long times) this
response is possibly non-symmetrical
with respect to the sign of the temperature changes,
it is still symmetrical in the relatively short times involved in
most simulations.

The longer the waiting time during
which the field has been applied,
the slower the relaxation of the
remanent magnetization.
If we normalize the magnetization by its value at $t_w$, then
the decay is a function of $t/t_w$.
This is again reminiscent of what happens in spin glasses and other disordered
systems, except for the fact that the susceptibility is finite in those cases.

%%%%%%%%%%%%%%%%%%%%%%%%%%%%%%%%%%%%%%%%%%%%%%%%%%%%%%%%%%%%%%%%%%%%%%%%

\vspace{1cm}
\section{Spinoidal Decomposition}
\setcounter{equation}{0}
\renewcommand{\theequation}{\thesection.\arabic{equation}}
\vspace{1cm}

%%%%%%%%%%%%%%%%%%%%%%%%%%%%%%%%%%%%%%%%%%%%%%%%%%%%%%%%%%%%%%%%%%%%%%%%

We consider a normal ferromagnetic system (of Ising or Heisenberg
type) and we
supose that the dynamics is local, without local conservation of the
magnetization. The Langevin equation described in the previous sections
is a
good example of such a dynamics.  For definiteness we  consider the
Ising case.

We are interested in studying the evolution of such a system when we
quench it
from high temperature to a subcritical temperature. The problem
is well
studied in the literature \cite{Br}. The main result is the random
formation of domains oriented in different directions which become
larger and
larger with increasing time. The size of the domains $\xi(t)$
grows
as $t^{1/2}$. It is also
well known that the equal time correlation function in
the large time limit is well described by
\beq
\la \phi(x,t) \phi(0,t) \ra = F(x/\xi(t)),
\eeq
for well separated space points; {\it i.e.} $x >> 1$.
The function $F$ is not very far from a Gaussian.

Throughout this section
the brackets stand for average over initial conditions.
We  suppose that the field $\phi$ at time zero is Gaussian-distributed
with a correlation function that goes to zero at large distances.

The most natural proposal for the correlation
function at different times is
 \beq
\la \phi(x,t') \phi(0,t) \ra = C(x/\xi(t),t/t')
\eeq
and, if we consider the correlation at the same space
point we would then have
\beq
\la \phi(0,t') \phi(0,t) \ra =C(t'/t)
\; .
\eeq

Intuitively we can understand this scaling as follows. The correlation
function is proportional to the probability that both spins stay in
the
same cluster. At time $t$ the spin stays in a cluster of size $\xi(t)$
which has a mean life proportional to $\xi(t)^2$, {\it i.e.} to $t$.
Therefore
it takes a time of order $t$ to revert the magnetization.

Let us be more precise. We consider the following zero temperature
Langevin
equation:
\beq
\frac{\partial}{\partial t} \phi
=  \Delta \phi +\phi (1 - \phi^2)
\; .
\label{langzero}
\eeq

One can treat this problem approximately as follows \cite{Ka}:
introduce a field
$\psi$ defined by
\beq
\phi=g(\psi)=\frac{\psi}{\sqrt{1+\psi^2}}
\label{phi}
\eeq
Eq. (\ref{langzero}) becomes, in terms of $\psi$,
\beq
\frac{\partial \psi }{\partial t}
=  \Delta \psi
+\psi
- \left( \frac{ \partial^2 g}{\partial \psi^2} \right)
\left( \frac{\partial g}{\partial \psi} \right)^{-1}
\left( \nabla \psi \right)^2
\; .
\label{languno}
\eeq
The approximation consists in neglecting the last term;
then one assumes
\beq
\frac{\partial \psi}{\partial t}
= \Delta \psi
+\psi
\; .
\eeq
Note the sign of the mass term.
We do not discuss here the range of validity of this
approximation,
which is widely done in the literature \cite{Br}.

The strategy we follow is similar to the one
used in the preceding section: we first solve a simple
linear problem and then calculate the physical correlations
as correlations of composite operators.

The $\psi$-correlation function in
Fourier space is given by
\beq
\la \psi({\bf k},t) \, \psi({\bf k'},t') \ra
\propto
\delta^D({\bf k} + {\bf k'})
\,
\exp( (t+t') (1-k^2))
\eeq
which corresponds to
\beq
\la \psi({\bf x},t) \, \psi({\bf x'},t') \ra
\propto
{1 \over (t+t')^{D/2}}\exp(t+t' -{x^2 \over 4(t+t')})
\eeq
in position space. For large times the absolute value of
$\psi$ becomes exponentially large, $\phi$
goes to $\pm 1$ ({\it cfr.}  eq. (\ref{phi})) and  one has
\beq
C(t,t')= \la \mbox{sgn}(\psi(0,t)) \, \mbox{sgn} (\psi(0,t'))\ra
\label{csign}
\; .
\eeq

The correlation of the random Gaussian
variables $\psi_1 \equiv \psi({\bf x}=0,t)$ and
$\psi_2 \equiv \psi({\bf x}=0,t')$,
is
\beq
\la \psi_i \, \psi_j \ra
\propto
{1 \over (t_i+t_j)^{D/2}} e^{t_i+t_j}
\;\;\;\;\;\;\;\;
i,j=1,2
\; ,
\eeq
their randomness comes from that of the initial conditions.

Eq. (\ref{csign}) becomes
\beq
C(t,t')
=
\frac{
\int d\psi_1 \, d\psi_2 \;
\mbox{sgn}(\psi_1) \mbox{sgn}(\psi_2)
\;
\exp
\left(
-\frac{\psi_1^2}{ 2\la \psi_1^2 \ra}
-\frac{\psi_1 \psi_2}{ \la \psi_1 \psi_2 \ra}
-\frac{\psi_2^2}{ 2\la \psi_2^2 \ra}
\right)
}
{
\int d\psi_1 \, d\psi_2 \;
\exp
\left(
-\frac{\psi_1^2}{ 2\la \psi_1^2 \ra}
-\frac{\psi_1 \psi_2}{ \la \psi_1 \psi_2 \ra}
-\frac{\psi_2^2}{ 2\la \psi_2^2 \ra}
\right)
}
\; .
\eeq
Changing variables
\beqa
\psi_1 &\rightarrow& \frac{\exp(t')}{(2t')^{D/4}} \; \psi_1 \nn
\psi_2  &\rightarrow& \frac{\exp(t')}{(2t')^{D/4}} \; \psi_2
\eeqa
we obtain
\beq
C(t,t')
=
\frac{
\int d\psi_1 \, d\psi_2 \;
\mbox{sgn}(\psi_1) \mbox{sgn}(\psi_2)
\;
\exp
\left(
-A \, \psi_1^2
- 2B \, \psi_1 \psi_2
-\psi_2^2
\right)
}
{
\int d\psi_1 \, d\psi_2 \;
\exp
\left(
-A \, \psi_1^2
-2 B \, \psi_1 \psi_2
-\psi_2^2
\right)
}
\; ,
\eeq
with
\beq
A=\lambda^{D/2} \;\;\;\;\;\;
B=\left(\frac{1+\lambda}{2}\right)^{D/2}
\; ,
\eeq
$\lambda = t'/t$.
Since we are only interested in the scaling we do not explicitly
compute this integral, but notice that
\beq
C(t,t') = C(\lambda)
\; .
\eeq

In a similar way one can prove the other scaling laws. The important
result
is that this approximation gives expressions for the correlations which
are in
very nice agreement with the aging formul\ae $\;$.

A similar analysis could have been done for the response function and
for the
correlation at finite temperature, but this would make this paper too
long.

\vspace{1cm}
\section{Conclusions}
\vspace{1cm}

We have seen that in many systems in which equilibrium is slowly
approached
some form of aging phenomena are present. These systems are characterized
by a
correlation
length that is infinite in the static limit but is finite for finite times:
it diverges with a power law in time.

A remarkable feature is that in these systems the energy
landscape is flat: no high barriers in energy are present. On the
contrary the flatness of the potential in certain directions, {\it i.e.} the
presence of zero modes, is at the origin of this very slow approach to
equilibrium.

These systems are an evident proof that it is not possible to conclude for
the existence of energy activated barrier-crossing only from the presence
of aging. It would be rather interesting to see if there are some peculiar
phenomena,  which
may distinguish
the effects of barriers from those due to flat directions.

%%%%%%%%%%%%%%%%%%%%%%%%%%%%%%%%%%%%%%%%%%%%%%%%%%%%%%%%%%%%%%

\vspace{1cm}
\noindent {\large {\bf {Appendix A}}}
\setcounter{equation}{0}
\renewcommand{\theequation}{A.\arabic{equation}}
\vspace{1cm}

%%%%%%%%%%%%%%%%%%%%%%%%%%%%%%%%%%%%%%%%%%%%%%%%%%%%%%%%%%%%%%

\noindent
The total response to a constant magnetic field $h\xr$ applied
during the interval $[0,t_w]$ over the whole system
defined in eq. (\ref{ther}) is
\beqa
\chi_{t_w}(t)
&=& \frac{1}{4}
(1+2t)^{-\frac{T}{8\pi}}
\int_o^{t_w} dt' \;
(1+2t')^{-\frac{T}{8\pi}}
\;\;
k \left( \frac{1+t-t'}{1+t+t'}, T \right)
\eeqa
with
\beq
k ( w, T )
\equiv
\int_o^\infty du \; \exp \left( -\frac{u}{2} \right)
\exp \left( -\frac{T}{4\pi}
\G
\left[
0; \frac{u}{2}, \frac{u w}{2}
\right]
\right)
\; .
\eeq
Changing variables
\beqa
\begin{array}{rclcrcl}
\omega
&=&
\frac{1+t-t'}{1+t+t'}
&
\;\;\;\;\;\;
\Rightarrow
\;\;\;\;\;\;
\omega_w
&=&
\frac{1+t-t_w}{1+t+t_w}
\end{array}
\eeqa
the total response is
\beq
\chi_{t_w}(t)
=
\frac{1}{2} (1+2t)^{-\frac{T}{8\pi}} (1+t)
\int_{\omega_w}^1
\frac{d\omega}{(1+\omega)^2}
\left[ 1 + 2 (1+t) \frac{1-\omega}{1+\omega} \right]^{-\frac{T}{8\pi}}
\;\; k(\omega,T)
\eeq
$\forall t,t'$.

Let us now consider the $\lambda$ scale, {\it i.e.} $t \gg  1$ and
$0 < \lambda_w \equiv t_w / t < 1$. In this case we can use
$1+a t \sim t$ and $1 + t-t' \sim t-t'$ in the lower limit
of the integral.
Then,
\beq
\chi_{t_w}(t)
=
2^{-(1+\frac{T}{4\pi})}
\;
\lambda_w^{-1+\frac{T}{4\pi}}
\;
t_w^{1-\frac{T}{4\pi}}
\int_{\frac{1-\lambda_w}{1+\lambda_w}}^1
\frac{d\omega}{(1+\omega)^2}
\left( \frac{1-\omega}{1+\omega} \right)^{-\frac{T}{8\pi}}
\;\;
k(\omega,T)
\label{formu}
\eeq
and finally
\beq
\chi_{t_w}(t)
=
t_w^{1-\frac{T}{4\pi}} \; f(\lambda_w,T)
\eeq
with $f(\lambda_w, t)$ given by eq.(\ref{funct}).
Note that the integrand in (\ref{formu}) diverges for $\omega=1$ and
gor $\omega=0$, but the integral over
$\omega$ is still convergent, and (\ref{formu}) is valid for all $\lambda_w$.

\vspace{3cm}

\noindent {\large {\bf Acknowledgements}}
\noindent This work originated in discussions
with Miguel Virasoro.
We are happy to thank him and Marc M\'ezard for useful discussions.

\pagebreak

\end{document}